
%
%
\input harvmac

\def\UCSD#1#2{\noindent#1\hfill #2%
\bigskip\supereject\global\hsize=\hsbody%
\footline={\hss\tenrm\folio\hss}}
\def\abstract#1{\centerline{\bf Abstract}\nobreak\medskip\nobreak\par #1}
\def\lamqcd{\Lambda_{\rm QCD}}
\def\frac#1#2{\textstyle{#1\over#2}}
\def\half{\frac12}
\def\mev{\,{\rm MeV}}
\def\gev{\,{\rm GeV}}

\noblackbox

\centerline{\titlefont{Recent Developments in the Theory}}
\bigskip
\centerline{\titlefont{of Heavy Quarks}}
\bigskip\bigskip
\centerline{Adam F.~Falk\footnote{\raise1ex\hbox{\hskip-.3em *}}{On leave from
The Johns Hopkins University, Baltimore, Maryland.}\footnote{}{To appear in the
Proceedings of the Advanced Study Conference on Heavy Flavours, Pavia, Italy,
September 3-7, 1993.}}
\bigskip
\vbox{\sl{\centerline{Department of Physics 0319}
\centerline{University of California, San Diego}
\centerline{9500 Gilman Drive, La Jolla CA 92093-0319}}}
\bigskip\vfill
\abstract{
Recent developments in the theory of heavy quarks are reviewed.
In the area of heavy quark fragmentation, there has been progress in the
study of both pertubative and nonperturbative processes, including the
identification of new observable nonperturbative fragmentation parameters.
There has also been considerable theoretical activity in the study of
inclusive rare and semileptonic decays of bottom hadrons.}
\bigskip
\UCSD{UCSD/PTH 93-31, JHU-TIPAC-930027}{October 1993}

\newsec{Introduction}

The properties of hadrons containing a heavy quark $Q$ simplify considerably
in the limit $m_Q\to\infty$.  If the
hadron contains only heavy quarks, this is because of asymptotic freedom.
If the hadron contains light quarks, this is because for $m_Q\gg\lamqcd$ the
light degrees of freedom become insensitive to the mass $m_Q$; as far as
they are concerned, the heavy quark acts simply as a non-recoiling source of
color.$^{1)}$  In the latter case, hyperfine effects associated with the
heavy quark chromomagnetic moment also decouple, and a new ``heavy quark
spin-flavor symmetry'' emerges.  There is an extensive literature in which
these simplifications have been used to predict hadronic spectra and decay
rates, among other things.  In this talk I will describe more recent work, in
which they have been applied to heavy quark fragmentation and inclusive
decays.

\newsec{Heavy Quark Fragmentation}

There has been much recent progress in describing the fragmentation to
hadrons of
heavy quarks which are produced at high energies in collisions.  In the case
of fragmentation to a hadron containing a single heavy quark, the symmetries
of the heavy quark effective theory have been applied to identify new
nonperturbative fragmentation parameters which describe the anisotropies of
the light degrees of freedom.  For fragmentation to hadrons containing two
heavy quarks, such as to ``onium'' states, perturbative QCD has been used
to compute fragmentation functions and probabilities in an
expansion in $\alpha_s(m_Q)$ and $m_Q/E$, where $E$ is the energy of the
quarkonium in the rest frame of the event.   We will discuss each of
these systems in turn.

\subsec{Heavy-Light Systems}

We begin by applying the heavy quark symmetries to fragmentation processes
in which no new heavy quarks are produced.$^{2)}$  In the limit $m_Q\to\infty$
such a process factorizes into short-distance and long-distance
pieces.  The heavy quark is first produced via some high energy interaction,
such as the decay of a virtual photon, which may typically be calculated in
perturbation theory.  This initial stage occurs on a time short compared to
the time
scale
$1/\lamqcd$ of the nonperturbative strong interactions.  Later, over a longer,
hadronic time scale, a fragmentation process occurs which eventually
produces a physical hadron containing the heavy quark.  Since this slower
process only involves the redistribution of energies small compared to $m_Q$,
the velocity of the heavy quark remains unchanged once it has been produced
perturbatively, and its mass and spin decouple from the nonperturbative
dynamics.

Let us now consider this sequence of events in somewhat more detail.
We imagine that we begin with a heavy quark which has been ejected at
relativistic speed from a hard reaction.  The axis linking the rest
frame of the heavy quark to the center of mass frame of the hard process
is a preferred direction, which we call the axis of fragmentation.
The interactions which couple to the heavy quark spin, such as the
chromomagnetic moment operator, are suppressed by at least one power of the
small parameter $\lamqcd/m_Q$, and hence the rate of heavy quark spin flip is
very slow on the hadronic time scale $1/\lamqcd$.  We might imagine the early
stages of fragmentation to involve the production of excited hadrons
containing the
heavy quark, which then rapidly decay to lighter excited states.  So long
as these decay times are shorter than the typical spin flip times
$m_Q/\lamqcd^2$, the heavy quark spin will be unaffected by this evolution.

However, eventually the light degrees of freedom will reach a state whose
lifetime is longer than the time required to flip the heavy quark spin.  In
the heavy quark limit, if the light degrees of freedom have angular momentum
$j>0$, the physical hadrons are a nearly degenerate doublet with spin
$J=j\pm\half$.  The trademark of a sufficiently long-lived state is that
the members of this doublet are well separated resonances; in that case, the
doublet lives long enough for the two eigenstates of total angular momentum
to become incoherent with each other.  This, in turn, will happen only if
the heavy quark and the light degrees of freedom have enough time to
exchange their spin orientations before the doublet decays.

It is at this point that things become interesting, since the dynamics now
allows the different helicity states of the light degrees of freedom (and of
the heavy quark) to mix with each other.  This effect, as we shall see below,
is observable in the subsequent decays of the long-lived states, in such
a way so as to give us information about the fragmentation process
itself.  We now introduce a new set of fragmentation
parameters appropriate to HQET.  When light degrees of freedom with angular
momentum $j$ are formed by fragmentation, they can be produced in one of
$2j+1$ helicity states along the fragmentation axis.  Parity invariance
requires
that the probability of forming a given helicity state cannot depend on the
sign of
the helicity $j^3$.  However, the fragmentation need not be entirely
isotropic; states with different magnitudes $|j^3|$ can arise with different
probabilities.  For the examples we will consider here, we can characterize
the situation as follows:  for a system of light degrees of freedom of
spin $j$, let $w_j$ be the probability that fragmentation leads to a state
with the maximum value of $j^3$.  Then the parameter $w_j$ may take values
between 0 and 1.

Of the systems available for study, the most interesting is the
$j={3\over2}$ excited meson doublet in the charm system.  These states have
been identified as the spin-1 $D_1(2420)$ and the spin-2 $D_2^*(2460)$, and
their decays to the ground state mesons $D$ and $D^*$ via pion emission
have been observed.$^{3)}$  The production of these resonances
via fragmentation is described by the parameter $w_{3/2}$, which is the
probability of producing the light degrees of freedom in a state of
helicity $\pm{3\over2}$ along the fragmentation axis, rather than
$\pm{1\over2}$.  Since the $D_1$ and $D_2^*$ peaks are well separated,
these two components of the doublet become incoherent before they decay,
and it becomes appropriate to describe them as a mixed state containing
$D_1$ or $D_2^*$ with fixed probabilities.  If we make the further
assumption that the initial charm quark is polarized left-handed (it is
straightforward to extend this to a general or mixed polarization), then the
populations of the helicity states of the excited mesons are easy to
calculate.  For example, the helicity states of the $D_2^*$ are populated with
the following relative probabilities, where the helicity runs from $-2$ to $+2$
across the table:
\eqn\helicities{
    p(D_2^*,h)=\bigg(\half w_{3/2} \qquad \frac38(1-w_{3/2}) \qquad
               \frac14(1-w_{3/2}) \qquad \frac18w_{3/2} \qquad 0\bigg)\,.}

Given these probabilities, we may now compute the angular distribution for
the observed strong decay $D_2^*\to D+\pi$.  We define $\theta$ to be
the angle between the momentum of the pion and the fragmentation axis, in
the rest frame of the $D_2^*$.  Normalizing to the full width $\Gamma$, we
find
\eqn\distrib{
    {1\over\Gamma}{d\Gamma\over d\cos\theta}(D_2^*\to D\pi)=
    \frac14\big[ 1+3\cos^2\theta-6w_{3/2}(\cos^2\theta-\frac13)\big]\,.}
This distribution is independent of the initial polarization state of the
charm quark (this is not necessarily the case for the decay of the $D_1$;
see Ref.~2.).  Note that for the isotropic production of the light
degrees of freedom, $w_{3/2}=\half$, the pion is also emitted
isotropically.  However, an ARGUS analysis of this decay$^{4)}$
found no significant population of the helicity states $\pm2$ of the
$D_2^*$, implying that $w_{3/2}$ is small.  Our own fit yields
\eqn\wfit{
    w_{3/2}<0.24\,, \qquad\qquad 90\%~{\rm conf.} }
This is an intriguing result, for which we have as yet no satisfactory
physical interpretation.

A similar quantity may soon be measured for the $j=1$ heavy charmed baryon
doublet, the spin-$\half$ $\Sigma_c$ and the spin-$\frac32$
$\Sigma_c^*$.  The $\Sigma_c$ has been observed with a mass of $2453\mev$;
while the $\Sigma_c^*$ has not yet been found, its mass is known to be
greater than
$2530\mev$.  In any case,
the peaks for the two states are known to be well separated.$^{3)}$  Hence an
analysis of the fragmentation parameter $w_1$ relevant to this system may be
performed by observing the strong decay $\Sigma_c^*\to\Lambda_c+\pi$.
Defining $\theta$ as before, we find the angular distribution of the pion to be
\eqn\bary{
    {1\over\Gamma}{d\Gamma\over d\cos\theta}(\Sigma_c^*\to\Lambda_c\pi)=
    \frac14\big[1+3\cos^2\theta-
    \frac92w_1(\cos^2\theta-\frac13)\big]\,.}
In this case, isotropic production of the light degrees of freedom
corresponds to $w_1=\frac23$.

This quantity, when it is obtained, will be useful for another reason.  The
ground state $\Lambda_Q$ baryon has light degrees of freedom with total
angular momentum $j=0$; the spin of this baryon is carried entirely by the
heavy quark.  Hence if a sample of $\Lambda_Q$ baryons arises from an
initial sample of polarized heavy quarks, they should themselves be strongly
polarized.$^{5)}$  (We have in mind charm and bottom quarks
produced at the $Z_0$ peak, which have polarizations $P_c=0.67$ and $P_b=0.94$,
respectively.)   The $\Lambda_Q$ would, in fact, carry all of the polarization
of the heavy quark if it were produced only directly, but there is a
potentially significant pollution from $\Lambda_Q$'s produced via the cascade
decays $(\Sigma_Q,\Sigma_Q^*)\to\Lambda_Q+\pi$.  If (as is the case in the
charm system) the excited doublet $(\Sigma_Q,\Sigma_Q^*)$ lives long enough
that spin exchange interactions between the heavy quark and the light degrees
of freedom have time to take place, the initial polarization of the heavy quark
will be diluted.  The actual polarization of the $\Lambda_Q$ baryons may
be computed in terms of the anisotropy parameter $w_1$ and a parameter $A$
which describes the  probability of producing a $j=1$ (and isospin one) diquark
relative to one with $j=0$ (and isospin zero), summed over the nine possible
helicity and isospin states.  If $P$ is the initial polarization of the heavy
quark, we find the polarization of the sample of $\Lambda_Q$'s to be
\eqn\lampol{
    P_\Lambda={1+(1+4w_1)A/9\over1+A}\,P\,.}
The parameter $A$ is related to one which appears in the Lund Monte
Carlo$^{6)}$
and may be estimated to be 0.45, with large
errors.  If, motivated by the results for charmed mesons, we take $w_1=0$,
we find a polarization $P_\Lambda=0.72 P$.  Note that the polarization
retention $P_\Lambda$ grows with $w_1$.  Finally, the brief treatment given
here
depends on the assumption that the $\Sigma_Q$ and $\Sigma_Q^*$ resonances
are well
separated.  While this is known to be true in the case of charm, it may not
necessarily be true for bottom.  For a more general analysis in which this
latter possibility is taken into account, see Ref.~2.

\subsec{Heavy-Heavy Systems}

We now turn to the much rarer possibility that the fragmentation process
generates an additional pair of heavy quarks, rather than just light
degrees of freedom.  Braaten {\it et al.}$^{7)}$ have
shown that the production of heavy ``onium'' states at high energies is
dominated by
such fragmentation processes, which may be calculated perturbatively.  A
typical process would be the production of charmonium from the
fragmentation of a high energy charm quark produced in $Z_0$ decay,
$Z_0\to \bar c c\to \psi + \bar c c + X$.  Braaten {\it et al.}~have shown that
such a process factorizes to leading order in $m_c/E$, where $E$ is the
energy of the $\psi$:
\eqn\factor{
    d\,\Gamma(Z^0\to\psi(E)+X)=\sum_i\int_0^1 dz\;
    d\,\hat\Gamma(Z_0\to i(E/z)+X,\mu)\,D_{i\to\psi}(z,\mu)\,.}
Here $d\,\hat\Gamma$ is the cross-section for the $Z_0$ to produce the
parton $i$ with energy $E/z$, and $D_{i\to\psi}(z,\mu)$ is a
process-independent fragmentation function for the parton $i$ to fragment
to a $\psi$ with momentum fraction $z$.  The scale $\mu$ is a factorization
scale which is introduced to separate the short- and long-distance physics;
note that the energy of the $\psi$ does not appear in
$D_{i\to\psi}(z,\mu)$, which may be evaluated at a low-energy scale
$\mu\approx m_c$.  The renormalization group may then be used to resum the
leading logarithms between $m_c$ and the high-energy scale $E$.

The fragmentation functions may be evaluated straightforwardly and
perturbatively in $\alpha_s(m_\psi)$, but the results are somewhat
messy.$^{7-9)}$   Of more interest are the fragmentation probabilites, obtained
by integrating over $z$.  In many cases it can be shown that the probabilites
$P=\int_0^1 dz\, D(z,\mu)$ are in fact independent of $\mu$, at least to
leading order.  Some typical probabilities which Braaten {\it et al.}~report
are\footnote{*}{I am grateful to T.~C.~Yuan for providing me with
updated versions of these numbers.}
\eqn\braaten{
    \eqalign{&P_{c\to\psi}\approx 1.2\times 10^{-4}\,,\cr
    &P_{b\to B_c}\approx 3.8\times 10^{-4}\,,\cr}\qquad\qquad
    \eqalign{&P_{c\to\eta_c}\approx 1.2\times 10^{-4}\,,\cr
     &P_{b\to B_c^*}\approx 5.4\times 10^{-4}\,.\cr}}

These results may be extended in a number of ways.  First, one may compute
not only the probability of a charm quark to fragment to a $\psi$, but the
fraction of the time that the $\psi$ is transversely rather than
longitudinally aligned.$^{8)}$  We define $\zeta$ to be
the ratio of the probability of producing a transversely aligned $\psi$ to
the total production probability.  Then a straightforward perturbative
calculation gives $\zeta=0.69$, corresponding to a small excess of
transversely aligned $\psi$'s. (This fraction is independent of the heavy
quark mass, and is hence the same for $\Upsilon$'s.)  In leading logarithmic
approximation, the corresponding ratio for gluon fragmentation to $\psi$'s
has the same value and is also $\mu$-independent.  Hence, at a hadron collider,
where $\psi$'s are produced by both quark and gluon fragmentation, the
fraction of transversely aligned $\psi$'s is also $\zeta=0.69$.

This fact has an immediate application to the study of direct $\psi$
production at hadron colliders, where it is important to distinguish these
$\psi$'s from those produced via the weak decays of $b$ quarks.  The alignment
of the $\psi$ may be observed in the angular distribution of its leptonic
decay, $\psi\to\ell^+\ell^-$, which is parameterized as
\eqn\leptons{
    {d\Gamma\over d\cos\theta}\propto 1+\alpha\cos^2\theta\,.}
Here $\theta$ is the angle between either of the leptons and the alignment
axis in the $\psi$ rest frame.  For $\zeta=0.69$, we find $\alpha=0.053$, a
small (5\%) asymmetry.  By contrast, for the weak decay $b\to\psi
s\ell^+\ell^-$ we find $\alpha\approx-0.46$.  Hence the lepton angular
distribution may provide a useful tool for separating these two sources of
$\psi$'s in a collider environment.

Finally, we have applied these perturbative methods to estimating the
probability of fragmentation to baryons which contain two heavy quarks,
such as $bbq$, $ccq$ and $bcq$ states.$^{9)}$  There are a
host of possible ``doubly heavy'' spin-$\half$ and spin-$\frac32$ baryons; we
found the largest production probabilities to be
\eqn\twoheavy{\eqalign{
    &P(c\to\Sigma_{cc},\Sigma^*_{cc})\sim 2\times 10^{-5}\,,\cr
    &P(b\to\Sigma_{bc},\Sigma^*_{bc})\sim 5\times 10^{-5}\,,\cr
    &P(b\to\Lambda_{bc})\sim 4\times 10^{-5}\,.\cr}}
While there is considerable uncertainty in these estimates (they should
be trusted at best within a factor of two), this is
enough to tantalize us with the hope that these states may one day be observed
at hadron colliders such as an upgraded Tevatron.

\newsec{Inclusive Heavy Hadron Decays}

There has been much recent progress in the analysis of inclusive decays of
heavy hadrons.  More precisely, these are decays to final states in which
some, but not all, of the quantum numbers of the decay products are known.
For example, we might specify the identity and momenta of certain weakly
interacting particles, such as leptons or photons, or we might restrict
the flavor of the final hadrons.  The decays are inclusive in that
we sum over all final states which can be produced by the long-distance,
nonperturbative strong interactions, subject to certain constraints which are
determined by short-distance, perturbative physics.

We are typically interested in studying quark-level transitions induced by
short-distance weak interactions, such as $b\to c\,\ell\bar\nu$, $b\to
s\gamma$, or $b\to se^+e^-$.  In a lagrangian renormalized far below the
weak scale, these interactions appear as local non-renormalizable operators
with coefficients which contain all dependence on the interesting
short-distance physics.  For example, the low-energy lagrangian contains
terms
\eqn\slop{
    {G_F\over\sqrt2}V_{cb} \bar
    c\gamma_\mu(1-\gamma^5)b\,\bar\ell\gamma^\mu(1-\gamma^5)\nu} and
\eqn\bsop{
    {4G_F\over\sqrt2}V_{tb}V_{ts}^*{e\over16\pi^2}m_bc_7(\mu)\,\bar
    s_R\sigma_{\mu\nu}b_L\,F^{\mu\nu}\,,}
and we would like to extract the parameters $V_{cb}$ and
$c_7(m_b)$ from experiment.  In order
to do this, however, it is necessary to understand how these quark-level
operators induce transitions in the physical hadrons which are actually
observed.

The difficulty is that exclusive decays, such as $B\to K^*\gamma$ in the
case of an underlying $b\to s\gamma$ transition, are typically not accessible
theoretically.  This is because they require the calculation of
nonperturbative hadronic matrix elements such as $\langle K^*|\,\bar
s_R\sigma_{\mu\nu}b_L\,|B\rangle$, for which there are no rigorous techniques
except in certain situations of enhanced symmetry.  It is simpler, in a
sense, to consider inclusive decays, in which all possble final states are
summed over.  One may constrain the kinematics by fixing the momenta of the
nonhadronic quanta in the final state, and one may assume in the sum that
the flavor of the final hadron may be tagged experimentally.  In that case,
what one needs instead of $\langle K^*|\,\bar
s_R\sigma_{\mu\nu}b_L\,|B\rangle$ is the sum over {\it all} strange hadrons of
the probabilities of producing them, \eqn\bssum{
     W=\sum_{X_s}\langle B|\,\bar b_L\sigma_{\alpha\beta}s_R\,|X_s\rangle
     \langle X_s|\,\bar s_R\sigma_{\mu\nu}b_L\,|B\rangle\,.}
There is an analogous expression, of course, for inclusive semileptonic
weak $b$ decays, and so forth.

If the energy which is released in the decay is large, then the inclusive
rate may be modeled simply by the decay of a free $b$ quark.  The
heuristic argument for this is simply that the quark-level transition takes
place over a time scale which is much shorter than the time it takes for the
quarks in the final state to materialize as physical hadrons.  Hence these two
parts of the hadronic decay process do not interfere with each other.  Once the
short-distance interaction has taken place, the probability is unity that
the quarks which have been produced will hadronize {\it somehow}; we do not
need to know the (incalculable) probabilities of hadronization to
individual final states if we sum over all of them.  This free quark decay
model (FQDM) has been the basis, until recently, of all theoretical studies
of inclusive heavy quark decays.

It is intuitively clear that the FQDM will be a good approximation when the
energy release in the decay is much larger than $\lamqcd$, and when the final
quark is far from its mass shell after the short-distance interactions have
taken place.  It has been shown by Chay {\it et al.}$^{10)}$ that
the FQDM may in fact be justified more rigorously within QCD, via an
operator product expansion.  To do this, one notes that the optical theorem may
be used to rewrite an inclusive sum such as \bssum\ as the imaginary part of a
forward scattering amplitude,
\eqn\top{
    W=2\,{\rm Im}\,T=2\,{\rm Im}\,\langle B|\,T\big\{\bar
    b_L\sigma^{\alpha\beta}s_R,
    \bar s_R\sigma^{\mu\nu}b_L\big\}\,|B\rangle\,.}
There is now no reference in the expression to any explicit strange
hadronic state
$X_s$.  Let us denote by $P^\mu$ the momentum of the virtual strange quark
in the
time-ordered product.  Since the scale of $P^\mu$ is set by the
available energy $m_b$, over almost all of the Dalitz plot we have $P^2\gg
m_s^2$ and the strange quark is far from its mass shell.  (The point in
the Dalitz plot is fixed by the kinematics of the rest of the event, in
this case
the energy of the photon.  For semileptonic decays, it would be set by the
lepton
kinematic invariants.)  If we stay away from those regions where
$P^2\approx m_s^2$, it is appropriate to expand the nonlocal hadronic
object $T$ in an operator product expansion, as a series of local operators
suppressed by the large off-shell momentum.

The insight of Chay {\it et al.}~was to notice that it is more convenient to
expand directly in inverse powers of the large mass $m_b$.  This is almost
the same thing as expanding in the off-shellness of the strange quark,
since $P^\mu$ scales with $m_b$; it only differs near those extreme regions
of the Dalitz plot where the expansion breaks down.  But this technique is
considerably more powerful, because it allow us to expand in operators of
the heavy quark effective theory$^{1)}$ (HQET).  The expansion of $T$ then has
the structure
\eqn\expansion{
    T\to {1\over m_b}\left[\CO_0+{1\over2m_b}\CO_1+{1\over4m_b^2}\CO_2
    +\cdots\right]\,,}
where the operator $\CO_n$ is of dimension $3+n$ and in HQET takes the form
\eqn\hqetops{
    \CO_n=c_n\bar h\Gamma D_{\mu_1}\ldots D_{\mu_n} h\,.}
The effective fields $h(x)$ which have replaced the quark fields $b(x)$
are obtained by the usual HQET rescaling.$^{1)}$

The advantage of the HQET formulation is that the heavy quark symmetries
may be used to compute the necessary matrix elements, up to the desired
order in $1/m_b$.  At leading order in the expansion of $T$, the matrix element
which is needed is of the form
\eqn\leading{
    \langle B|\,\bar h\Gamma h\,|B\rangle\,.}
This may be computed exactly in HQET; it is proportional to the Isgur-Wise
function evaluated at the zero-recoil point, where it is
absolutely normalized.  Chay {\it et al.}~pointed out that the result obtained
by truncating $T$ at this leading operator reproduced the free quark decay
model prediction for the inclusive rate.  Hence the FQDM was shown to be the
first term in a controlled expansion in $m_b$, with all corrections formally
of order $\lamqcd/m_b$.  This was a considerable improvement from its earlier
status as an {\it ad hoc}, if intuitively reasonable, model.

In fact, Chay {\it et al.}~showed that the situation is considerably better
than
it appears at first glance, because all corrections of relative order
$\lamqcd/m_b$
vanish if the decay rate is written in terms of the bottom {\it quark}, rather
than {\it meson}, mass.  (The distinction is significant, since the
decay rate is proportional to $m_b^5$.)  The result follows simply from the
fact that all matrix elements of HQET operators with one derivative vanish at
the point of zero momentum transfer.$^{11)}$  This is a surprising and
intriguing result, since there are certainly $1/m_b$ corrections to the
individual exclusive modes; however in the inclusive sum they conspire
to cancel.

The recent progress in the field has arisen due to the realization that the
leading non-vanishing corrections, of relative order $1/m_b^2$, may be
computed simply in terms of two nonperturbative parameters.  All the
necessary matrix elements of operators $\CO_2$ may be expressed in terms of
the matrix elements of two dimension-five operators$^{12)}$:
\eqn\dimfive{\eqalign{
    &\langle B|\,\bar h(iD)^2h\,|B\rangle=2m_B\lambda_1\,,\cr
    &\langle B|\,\bar h(-\frac{i}2\sigma^{\mu\nu})G_{\mu\nu}h\,|B\rangle
    =6M_B\lambda_2\,.\cr}}
The parameters $\lambda_1$ and $\lambda_2$ appear also in the expansion of
the physical hadron masses in terms of the bare quark mass,
\eqn\masses{\eqalign{
    &M_B=m_b+\bar\Lambda-{1\over2m_b}(\lambda_1+3\lambda_2)\,,\cr
    &M_{B^*}=m_b+\bar\Lambda-{1\over2m_b}(\lambda_1-\lambda_2)\,.\cr}}
Here $\bar\Lambda$ is the energy of the light degrees of freedom.$^{11,13)}$
The physical splitting between the vector and pseudoscalar states may be
used to extract $\lambda_2$, which represents the additional energy of the
$b$ quark in the hadron due to its chromomagnetic interactions with the
light degrees of freedom.  We find $\lambda_2=0.12\gev^2$.  Unfortunately,
$\lambda_1$, which represents the
$b$ quark's residual kinetic energy in the bound state, is not easily
related to any such observable.  It is probably reasonable to allow it to vary
within the range $-1\gev^2<\lambda_1<1\gev^2$.

We can now obtain the leading nonperturbative corrections to the inclusive
width $\Gamma$, as well as to other inclusive observables.  The expansion will
generally take the form
\eqn\gamexand{
    \Gamma\propto m_b^5\left[1+{1\over4m_b^2}f(\lambda_1,\lambda_2)
    +\cdots\right]\,.}
Here the ellipses denote corrections arising from yet higher order terms in
the operator product expansion, plus QCD radiative corrections which we have
not included.  We stress that while such
radiative effects, such as real and virtual gluon emission, can be as
large as the terms which we consider here, they may simply be added
in as necessary.

We close by tabulating some recent results which have been obtained with
these methods.  The most extensive work has concerned inclusive
semileptonic $b\to c$ transitions.$^{14)}$  One may obtain
the leading corrections to the full inclusive width, as well as corrections
to the lepton energy spectrum.  Since Prof.~Bigi has discussed his own
results at this conference, I have little to add.  The corrections to the
lepton spectrum are typically a few percent, except near the endpoint of
maximum energy.  There the effect grows dramatically, but these divergences
are a signal of the breakdown of the operator product expansion and are not
to be trusted.  The computed spectrum must be smeared over lepton energies
of a few hundreds of MeV for the results to be sensible.  Unfortunately it
is not possible to probe the very endpoint of the lepton energy spectrum in
detail using these techniques.  Hence the study of the charmless
semileptonic transition $b\to u\ell\bar\nu$ is not practical.

These techniques have also been applied to the rare $b\to s$ transitions
$B\to X_s\gamma$ and $B\to X_s\ell^+\ell^-$.  For the two-body final state,
we find corrections to the inclusive width,$^{15,16)}$
\eqn\gammcorr{
    \Gamma\propto m_b^5\left[1+{1\over2m_b^2}(\lambda_1-9\lambda_2)\right]
    \,,}
and to the average photon energy,$^{15)}$
\eqn\enave{
    \langle E_\gamma \rangle = {m_b\over2}\left[1-{1\over2m_b^2}
    (\lambda_1+3\lambda_2)\right]\,.}
Note that it is the quark mass $m_b$ which appears in these expressions,
not the meson mass.  The corrections are typically a few percent, which
tells us that the FQDM is in fact a good approximation when applied to
these rare $b$ decays.  Finally, we observe that the form of the correction
to $\langle E_\gamma\rangle$ is just what we would expect, since,
according to eq.~\masses,
$-(\lambda_1+3\lambda_2)/2m_b$ is precisely the shift in the energy of the
free $b$ quark when it is bound in the $B$ meson.

For the $X_s\ell^+\ell^-$ final state, the form of the result is too
unwieldy to reproduce here, and we refer the reader to Ref.~15 for details.  We
have calculated the leading corrections to the lepton invariant mass spectrum,
$d\Gamma/d\hat s$, where $\hat s=(P_{\ell^+}+ P_{\ell^-})^2/m_b^2$.  They are
somewhat larger than for $b\to s\gamma$, typically on the order of ten
percent.  However, one must be careful to stay away from the $\psi$ resonance
region in $\hat s$, where additional four-quark operators can contribute and
the operator product expansion is not to be trusted.

\bigskip\bigskip

{\bf Acknowledgements}
\medskip

It is a pleasure to thank the organizers of the conference for their warm
hospitality and flawless organization.  This work was
supported by the Department of Energy under contract
DE-FG03-90ER40546 and by the Texas National Laboratory Research
Commission under grant RGFY93-206.

\bigskip\bigskip

{\bf References}
\medskip
\frenchspacing

1.~N.~Isgur and M.B.~Wise, {\sl Phys.~Lett.}~{\bf B232} (1989) 113; {\bf
B237} (1990) 527;  M.B.~Voloshin and M.A.~Shifman, {\sl Yad.~Fiz.}~{\bf 45}
(1987)
463 [{\sl Sov.~J.~Nucl.~Phys.} {\bf 45} (1987) 292]; H.D.~Politzer and
M.B.~Wise,
{\sl Phys.~Lett.}~{\bf B206} (1988) 681; {\bf B208} (1988) 504; E.~Eichten and
B.~Hill, {\sl Phys.~Lett.}~{\bf B234} (1990) 511; B.~Grinstein,
{\sl Nucl.~Phys.}~{\bf B339} (1990) 253.

2.~A.F.~Falk and M.E.~Peskin, SLAC-PUB-6311, August 1993.

3.~K.~Hikasa {\it et al.} (Particle Data Group), {\sl Phys.~Rev.}~{\bf D45}
(1992), No. 11-II.

4.~H.~Albrecht {\it et al.} (ARGUS collaboration), {\sl Phys.~Lett.}~{\bf B221}
(1989).

5.~T.~Mannel and G.A.~Schuler, {\sl Phys.~Lett.}~{\bf B279} (1992) 194;
F.E.~Close, J.~K\"orner, R.J.N.~Phillips and D.J.~Summers, {\sl
J.~Phys.}~{\bf G18}
(1992) 1716.

6.~T. Sj\"ostrand, {\sl Comp.~Phys.~Comm.}~{\bf 39} (1986) 347.

7. E. Braaten and T.C. Yuan, NUHEP-92-93 (1992);
E. Braaten, K. Cheung and T.C. Yuan, NUHEP-TH-93-2 (1993);
E. Braaten and T.C. Yuan, NUHEP-TH-93-6 (1993).

8. A.F. Falk, M. Luke, M.J. Savage and M.B. Wise, {\sl Phys. Lett.} {\bf B312}
(1993) 486.

9. A.F. Falk, M. Luke, M.J. Savage and M.B. Wise, {\sl Phys. Rev.} {\bf D},
Brief
Reports, in press.

10. J. Chay, H. Georgi and B. Grinstein, {\sl Phys. Lett.} {\bf B247}
(1990) 399.

11. M. Luke, {\sl Phys. Lett.} {\bf B252} (1990) 447.

12. A.F. Falk and M. Neubert, {\sl Phys. Rev.} {\bf D47} (1993) 2965, 2982.

13. A.F. Falk, M. Luke and M. Neubert, {\sl Nucl. Phys.} {\bf B388} (1992) 363.

14. I.I. Bigi, M. Shifman, N.G. Ulratsev and A.I. Vainshtein, {\sl Phys. Rev.
Lett.} {\bf 71} (1993) 496; B. Blok, L. Koyrakh, M. Shifman and A.I.
Vainshtein,
NSF-ITP-93-68 (1993); A. Manohar and M.B. Wise, UCSD/PTH 93-14 (1993); T.
Mannel,
IKDA 93/26 (1993).

15. A.F. Falk, M. Luke and M.J. Savage, UCSD/PTH 93-23, August 1993.

16. I.I. Bigi, N.G. Uraltsev and A.I. Vainshtein, {\sl Phys. Lett.} {\bf B293}
(1992) 430; I.I. Bigi, B. Blok, M. Shifman, N.G. Uraltsev and A.I. Vainshtein,
TPI-MINN-92/67-T (1992).

\end